# Integrated Broadband Mode Division Demultiplexer in Waveguide Arrays

Yu-le Zhao, Chong Sheng, Zi-yi Liu, Shi-ning Zhu, Hui Liu*

*National Laboratory of Solid State Microstructures, School of Physics, Collaborative Innovation Center of Advanced Microstructures, Nanjing University, Nanjing 210093, China*

*liuhui@nju.edu.cn

**ABSTRACT**

On-chip mode division (de)multiplexing plays a significant role in the integrated devices to greatly improve communication capacity. One of the pursuing goals of mode division (de)multiplexer is how to realize such devices with larger bandwidth and better fabrication tolerance. However, integrated broadband mode division (de)multiplexing devices were rarely reported up to now. In this work, we experimentally realize a broadband mode division demultiplexing device with a high degree of fabrication tolerance based on lithium niobate-on-insulator waveguide array. By taking advantage of the fact that different modes in the waveguide array have different group velocity, we experimentally confirmed that different modes in the waveguide array have different transverse displacements. The implementation of the integrated broadband mode division demultiplexer may have future applications in information processing technology, quantum communication as well as quantum computing.

## 1. INTRODUCTION

With the exponential growth of bandwidth demand for on-chip communication and optical fiber communication systems, one of the urgent problems to be solved is expanding the transmission capacity in integrated optical communication systems. To improve the transmission capacity of communication, one can apply wavelength division multiplexing technology [1-3] and space division multiplexing technology [4-6] into communication systems, which provide a high degree of freedom for carrying information. Furthermore, recent years also have witnessed remarkable advances in mode division multiplexing technology, which uses the freedom of optical orthogonal mode of a single wavelength to increase the optical communication capacity. In particular, this technology can be deployed together with wavelength division multiplexing and polarization multiplexing, so that the communication capacity can be further increased many times. Various types of mode division (de)multiplexing devices have been achieved, including asymmetric Y-junction type [7-9], adiabatic coupling type [10], inverse grating coupling type [11-13], asymmetric directional coupling type [14-18], multimode interference type [19], and inversely designed structures [20,21]. Nevertheless, the design of the asymmetric Y junction requires a relatively large footprint which is not favor in the integration. And the inverse grating coupling type with the subwavelength structure requires a high-precision fabricate technology but has limited operation bandwidth. Additionally, the asymmetric directional coupling type not only has a large occupancy area but also requires the accurate fabrication of the designed structures, which also has the disadvantage of narrow working bandwidth. While some research groups have utilized the method of reverse design to realize an ultra-compact footprint and improve the working bandwidth, but the reverse design requires a large amount of calculation and accurate sample fabrication technology. And the adiabatic coupler has the advantages of easy design and fabrication among these mode division (de)multiplexing devices, but only supports a few optical modes. To date, most mode division (de)multiplexers are either wavelength sensitive or need a high requirement of fabrication. Therefore, how to realize integrated mode division (de)multiplexing devices with broadband working wavelength and easy fabrication has attracted a large amount of attention.

At the same time, Lithium niobate on insulator (LNOI) has strong electro-optic, acousto-optic, and thermo-optic effects with the wide transparent window of the wavelength as well as a high refractive index contrast and a nonlinear coefficient, so that it has become a promising candidate for integrated photonics. In fact, some basic structures have been developed on the LNOI platform [22-24], such as waveguides [25-27], and micro-ring resonators [28,29]. Moreover, a number of integrated devices have also been developed on this platform, such as ultra-compact on-chip spectrometer [30], electro-optic modulators [31], and optical frequency combs [32]. However, the works about broadband mode division demultiplexing on the LNOI platform have rarely been reported to date.

In this work, we use multimode waveguide arrays to achieve broadband mode division demultiplexing functionality. This is mainly due to the fact that the group velocity for different modes in this designed structure is mode sensitive but not wavelength sensitive. To realize such devices, we first calculate group velocities of different modes in the Lithium Niobate Thin film waveguide array using COMSOL Multiphysics (COMSOL Inc.). Then we experimentally fabricated the designed lithium niobate waveguide arrays for the proof of concept of broadband mode division demultiplexers. To experimentally demonstrate mode division demultiplexing in waveguide arrays, several modes were input into the waveguide arrays through the multi-mode waveguide, and then the transmission paths of different modes were observed through fluorescence imaging. Our results clarified that different modes in the waveguide array have different transverse displacements. Additionally, to exhibit the broadband characteristics of this device, we employed different wavelengths in a certain bandwidth range in this waveguide arrays. As a result, this broadband mode division demultiplexer can increase the bandwidth density of the on-chip interconnection and improve the information transmission capacity of the integrated communication system.

## 2. DEVICE STRUCTURE AND OPERATION PRINCIPLE

To achieve the mode division demultiplexing in waveguide arrays, we first study multiple modes for a single waveguide, and then we study the mode coupling behavior in two identical waveguides. For a single waveguide as shown in Fig.1(a), the electromagnetic wave with a wavelength of 445 nm can support multiple modes transmitted within it. Fig.1(a) shows the mode field distribution of several TE modes, including $TE_0$, $TE_1$, and $TE_2$ mode. According to waveguide mode theory, each mode in a multimode waveguide has a different propagation constant and a different mode cross-sectional area. Furthermore, based on the mode-coupling theory, the same waveguide modes in the adjacent two identical multimode waveguides will form the symmetric and antisymmetric modes due to coupling, as shown in Fig.1(b)-(d). For different waveguide modes, we found that the coupling coefficient increased with the higher mode order. Moreover, the propagation constants of symmetric and antisymmetric modes generate disparity owing to the coupling between waveguides. Remarkably, the relation between the coupling coefficient and the disparity of symmetric and anti-symmetric modes can be described as $\kappa_m = (\beta_{m,s} - \beta_{m,a})/2$, $\beta_{m,s}(\beta_{m,a})$ is the propagation constant of the *m*th order symmetric (antisymmetric) mode, The values of $\beta_{m,s}$ and $\beta_{m,a}$ are numerically calculated in COMSOL. We also calculated the coupling coefficients by an analytical method that is calculating the mode field overlap integral of waveguide modes[33]. We compared the coupling coefficients calculated by the numerical method with the analytical method, and the results calculated by both are fitting well as shown in Fig.1 (e).

To study the evolution of different waveguide modes in a waveguide array, we need to analyze the mode dispersion caused by the coupling among waveguides. According to the coupled wave theory, the electromagnetic wave transmitted within the waveguide arrays satisfies the following equation [34] :

$$i\frac{da_{n,m}(z)}{dz} + \kappa_m[a_{n-1,m}(z) + a_{n+1,m}(z)] = 0 \qquad (1)$$

Here, $a_{n,m}(z)$ represents the amplitude of the *m*th order mode electric field in the *n*th waveguide. The dispersion relation of the *m*th order mode in the waveguide arrays can be depicted as $k_{z,m} = \beta_m + 2\kappa_m\cos(k_{x,m}D)$, $\beta_m$ is the propagation constant of the *m*th order mode in the single waveguide. $k_{x,m}(k_{z,m})$ is transverse (longitudinal) of the *m*th order mode wave vector. $D$ is the period of the waveguide arrays. According to this formula, the dispersion of the first three TE modes are obtained by an analytical method as shown in Fig. 2(a). In addition to obtaining the periodic dispersion by the analytical method, we also calculated the periodic dispersion by directly simulating the waveguide array in COMSOL. Fig. 2(a) shows that the dispersions obtained by the two methods are fitting well. Then we define the group velocity $V_{x,m}$ and $V_{z,m}$ for the *m*th order waveguide mode, which are respectively along the transverse direction and the longitudinal direction. We also define the mode division angle (MDA) of the *m*th order mode in the waveguide array as $\theta_m$, which is depicted as $\tan\theta_m = v_{x,m}/v_{z,m} = \frac{d\omega}{dk_{x,m}}/\frac{d\omega}{dk_{z,m}} = dk_{z,m}/dk_{x,m}$, thus $\theta_m = \arctan[dk_{z,m}/dk_{x,m}]$. Details of the numerical method to calculating MDA can be found in Section I of the Supplemental Material[35]. During mode transmission from a single waveguide to a waveguide array, the propagation constants of the *m*th order mode need to satisfy the conservation condition, which can be depicted as $k_{z,m} = \beta_m$, and $\beta_m$ is the propagation constant of the *m*th order mode in the single waveguide. Considering also $k_{z,m} = \beta_m + 2\kappa_m\cos(k_{x,m}D)$, we then know that $k_{x,m} = \frac{\pi}{2D}$. Finally, we can obtain the following formula:

$$\theta_m = \arctan(-2D\kappa_m) \qquad (2)$$

The above formula gives the calculation of MDA. From the formula, we can find that the larger the coupling coefficient (larger the mode order), the larger the MDA in the waveguide array will be, as shown in Fig. 2(b).

When considering the different wavelengths of the incident electromagnetic wave, we can also calculate the dispersion curve corresponding to the specific mode of each wavelength, as shown in Fig. 3 (a), $kz(kx,\lambda)$ of several modes under the change of wavelengths from 405 nm to 505 nm. In the surface diagram, the three surfaces from top to bottom represent $TE_0$, $TE_1$, and $TE_2$ in turn. The arrows of different colors on the dispersion curve indicate the different modes of MDA. To explore the broadband potential of the scheme, we need to analyze the variation of MDA with wavelength as illustrated in Fig. 3 (a). Therefore, we have calculated the curve of MDA with wavelength, as shown in Fig. 3 (b). The open label in Fig. 3 (b) is directly obtained from the experimental data, indicating the MDA of three TE modes when several wavelengths selected in the experimental test are excited. We found that the MDA of each mode increases with the increase of wavelength between 405 nm and 505 nm, and the MDA of the lower-order mode increases slower than that of the higher-order mode. When the excitation wavelength of $TE_2$ mode is changed from 405 nm to 505 nm, the MDA changes by about 4°. Consequently, the MDA increases slowly with the wavelength. It can be inferred that the designed waveguide array structure can meet the mode division demultiplexing in a relatively wide wavelength range, and can be used as a broadband mode demultiplexing device.

## 3. SAMPLE FABRICATION AND EXPERIMENTAL TESTING

The design principles described above for broadband mode division demultiplexing are applicable to many material platforms, including silicon on insulator (SOI) and silicon nitride, etc. In each material platform corresponding to the waveguide arrays system, the MDA of the corresponding mode can be calculated based on the group velocity theory. In this experiment, we fabricated the broadband mode division demultiplexer based on LNOI. First, we drilled the waveguide arrays with a focused ion beam, together with the couple-in tapers and gratings. The SEM of the lithium niobate waveguide arrays is shown in Fig. 4 (a-b), where the detailed structural parameters can be found in Section III of the Supplemental Material[36]. The grating coupler can couple the laser into a single multimode waveguide, where the excited $TE_0$,

$TE_1$, and $TE_2$ modes will be introduced into the waveguide arrays by the single multimode waveguide. In the second step, to directly observe the mode transmission path, we used rare earth fluorescence imaging technology. The europium-doped PMMA solution will emit fluorescence in the 610 nm band when it is irradiated by a 400-500 nm waveguide laser. The fluorescent material used here is a common commercial fluorescent material with a luminous wavelength of 610 nm that can be excited by shorter wavelengths. Indeed, if we can find a fluorescent material with a luminescence wavelength larger than 2 um, then we may utilize the communication band (near 1550 nm) for experimental testing. We prepared a PMMA solution containing europium, which was prepared by mixing PMMA particles with rare earth europium particles into a toluene solution. The detailed preparation process of the PMMA solution can be found in Section III of the Supplemental Material[37]. After preparing the waveguide arrays and europium doped PMMA solution, we spun the above solution onto the lithium niobate chip through a homogenizer and obtained a sample of the waveguide arrays that suit our experimental observation requirements[38]. Here, we chose the proper PMMA concentration and spin coating thickness so that we would get a good fluorescence effect.

To demonstrate broadband mode division demultiplexing, we utilized the samples prepared in the above process for experimental testing. In the experimental demonstration, we coupled 430 nm, 445 nm, 470 nm, and 480 nm lasers into the waveguide arrays through the grating coupler, respectively. Different modes will be transmitted in the waveguide array, and the mode transmission paths will be revealed by fluorescence imaging. Considering that we need to observe the 610 nm fluorescence image, we used a filter to filtering the short wavelength laser and a SCMOS camera to capture images of the mode transmission path. The images of the mode transmission paths at different wavelengths are shown in Fig. 4 (d), in which mode transmission paths at different angles represent different modes of transmission. Correspondingly, Fig.4 (c) shows the transmission schematic corresponding to TE modes excited at different wavelengths calculated by group velocity theory. It can be seen that the experimental results and theoretical prediction results under several wavelengths of excitation within a certain bandwidth are in good agreement. We extracted the average value of the mode angle obtained from the experimental test and compared it with the mode angle obtained from the theoretical calculation, as shown in Table 1. It follows that there is a slight deviation between the experimental test angle and the theoretical prediction angle, but the deviation is within a reasonable range. This discrepancy may arise from deviations in sample preparation. In addition, comparing the fluorescence trajectories excited at different wavelengths in Fig.4 (d), we can find that the mode energy percentage is different. On the one hand, this may be due to the different coupling efficiency of the gratings at different wavelengths; on the other hand, it may be caused by the tolerance of sample processing. Detailed experimental tests can be obtained in Section III of the supplementary materials[39]. Given the above, this experiment has realized the conceptual verification of broadband mode division demultiplexing based on LNOI.

## 4. DISCUSSION

In addition to focusing on the broadband characteristics of mode division demultiplexing, we are also interested in the fabricate tolerance property of waveguide arrays mode division demultiplexing devices. Therefore, we also theoretically calculated the influence of different sidewall inclination angles on the MDA. The results are shown in Fig.S1(b) in the supplementary material[40]. It is found that the sidewall inclination angle has little effect on the MDA. That is, the different sidewall inclination angles caused by fabricating factors have little effect on the MDA. As a result, the mode division demultiplexing realized by the lithium niobate waveguide arrays has good fabricate tolerance. To improve the efficiency and performance of the mode division demultiplexer, we further optimized the design by propagation simulation. In the propagation simulation, we defined the mode energy output port of the demultiplexer, while we calculate the mode demultiplexing efficiency. The results show that the mode demultiplexing efficiency is wavelength insensitive, which indicates that the device has the potential for broadband operation. Detailed propagation simulation results and analysis can be found in Appendix A through C. Fluorescence characterization allows us to directly see the transmission path of the light field, which makes it convenient for us to visually judge the transmission path of the mode. Moreover, another measurement technique has attracted our attention, i.e., imaging the output facet of the waveguide array. This measurement technique will offer cleaner data with more information. We will try to use this measurement method in our future experiments. The design principle of the mode division demultiplexer we use can theoretically support more modes. In addition, using the same operating principle, we can tune the device parameters to support higher order modes for mode division demultiplexing.

## 5. SUMMARY AND OUTLOOK

In this work, we experimentally achieved mode division demultiplexing by exploiting the group velocity difference of different modes in an LNOI waveguide array. We begin by theoretically analyzing waveguide arrays mode division demultiplexing and then conceptually verifying the broadband property of waveguide arrays mode division demultiplexing through experiments. Specifically, different modes have different transmission angles for the input direction of a single waveguide, which causes the output modes to be located at different output ports, thus realizing the path allocation of different modes. By adjusting the structure of a uniform waveguide array, such as the gap between waveguides and the width of waveguides, we can design a mode demultiplexer that meets our requirements. Moreover, the mode transmission angle is mode sensitive but not wavelength sensitive. Therefore, a uniform waveguide array with a certain structure has the function of a broadband mode division demultiplexer. On the one hand, LNOI has a high nonlinear coefficient and is a good candidate for preparing efficient quantum entangled light sources; On the other hand, encoding quantum signals in waveguide mode is conducive to improving the speed of quantum information processing. Therefore, the integrated broadband mode division demultiplexing realized on LNOI may pave the way for the quantum photonic integrated circuit-based quantum information process. Further, the non-uniform waveguide array has more adjustable parameters, such as waveguide width, the gap between waveguides, etc. Thus, the non-uniform waveguide array mode division demultiplexer has the opportunity to improve the device performance compared with the uniform waveguide array mode division demultiplexer. Typically, the strong second-order nonlinear of lithium niobate is friendly for the application of frequency down-conversion device. Therefore, with a proper waveguide array structure design, we can achieve mode division demultiplexing during frequency down-conversion, which has the opportunity to further increase the transmission capacity in integrated optical communication systems. At the same time, lithium niobate possesses a strong electro-optical effect and is a popular material platform for electro-optical modulators. Therefore, the integration of electro-optical modulators and mode division demultiplexers may lead to interesting physics, such as the implementation of switch modes post sorting, which is what we want to explore in the future.

**ACKNOWLEDGMENTS**

This work financially supported by the National Natural Science Foundation of China (Grant Nos. 92163216, 92150302, 62288101, 12174187) and the Fundamental Research Fund for the Central Universities, China (Grant No. 14380139, 14380191).

# APPENDIX A: THE SIMULATION OF MODE TRANSMISSION AND ITS COMPARISON WITH EXPERIMENTAL RESULTS

In this work, we utilized grating couplers to excite multiple modes. While we acknowledge that the excitation method may not be as precise as desired, and the energy ratio for different modes cannot be strictly controlled, our main objective is to demonstrate a unique approach for broad mode division demultiplexing with high efficiency. We note that precise control of mode division multiplexing is currently a hot topic of research, and various techniques have been employed in pursuit of this goal for more precise control of information transmission. In future work, we will explore the use of broadband couplers to couple modes of different orders into the waveguide array in hopes of improving control over the process.

Although we cannot yet control the energy ratio for different waveguide modes with a high degree of accuracy, we can confirm that different waveguide modes have been generated using grating coupling. This point is well-demonstrated through the three-dimensional (3D) finite-difference time domain (FDTD) simulation, as shown in Fig. 5(c), which depicts the propagation simulation of different waveguide modes under 445 nm excitation. To compare with the experimental results, we assume that the energy ratios of $TE_0$, $TE_1$, and $TE_2$ modes are 50%, 25%, and 25%, respectively. Using data matrix processing software, we obtain the propagation light field diagram shown in Fig. 5(a), which closely resembles the experimental result shown in Fig.5(b). This agreement between our experimental and simulation results strengthens our confidence in our observations and conclusions.

# APPENDIX B: OPTIMIZING THE MODE INPUT METHOD

It has come to our attention that existing mode demultiplexers incur an efficiency loss of 3 dB. To address this issue, we have explored the boundary waveguide excitation method. Specifically, we conducted propagation simulations using FDTD simulation, as depicted in Fig. 6(a) and (c). When exciting the $TE_2$ mode from the waveguide at the boundary, we observed that it was transmitted to the right side of the waveguide array, as shown in Fig. 6(c). This result contrasts with those obtained using middle waveguide excitation, as presented in Fig. 6(a). Additionally, Fig. 6(b) and Fig. 6(d) illustrate the mode distribution at the output facet for boundary and middle waveguide excitation, respectively. Furthermore, we calculated the mode demultiplexing efficiency for both excitation methods. We defined this efficiency as the ratio of output power in the defined output port (indicated by the red dashed box region in Fig. 6(b) and Fig. 6(d)) to the input power of the main input waveguide. Our results demonstrate that the mode demultiplexing efficiency for boundary and middle waveguide excitation is 29.1% and 88.3%, respectively.

**APPENDIX C: THE DESIGN OF OUTPUT PORTS AND THE CALCULATION OF DEMULTIPLEXING EFFICIENCY**

To further investigate the broadband properties of our device, we designed an output port in our propagation simulation, as illustrated in Fig. 7(a). This port is defined as the mode energy output port (MEOP) and comprises a waveguide directly connected to the main input waveguide, named Out0, for the $TE_0$ mode. The MEOP for the $TE_1$ mode is a tapered waveguide connecting with the 2nd to 5th waveguides, named Out1, while for the $TE_2$ mode, it is a tapered waveguide connecting with the 6th to 10th waveguides, named Out2. As demonstrated in Fig. 7(b-d), when using a 405 nm excitation wavelength, the power distribution of the $TE_0$ mode is mainly concentrated on the main waveguide. Conversely, the power distributions of the $TE_1$ and $TE_2$ modes are concentrated on the 2nd to 5th and 6th to 10th waveguides, respectively. This arrangement enables us to route the energy of each mode to the corresponding MEOP.

We calculated the mode demultiplexing efficiency for each mode at a certain bandwidth using the ratio of output power in the defined output port to the input power of the main input waveguide. As indicated in Fig. 7(e), when the excitation wavelength is varied from 405 nm to 505 nm, the mode demultiplexing efficiency of the $TE_0$, $TE_1$, and TE$_2$ modes varies slowly with wavelength. These results suggest that our designed mode division demultiplexer exhibits relatively high demultiplexing efficiency for each mode at a bandwidth of 100 nm. Although we have not yet carried out relevant experiments, our simulations indicate that our devices possess broadband operation capabilities. Therefore, we plan to experimentally route several different transverse waveguide modes to the different output ports of the waveguide array in future work.

**Figure**

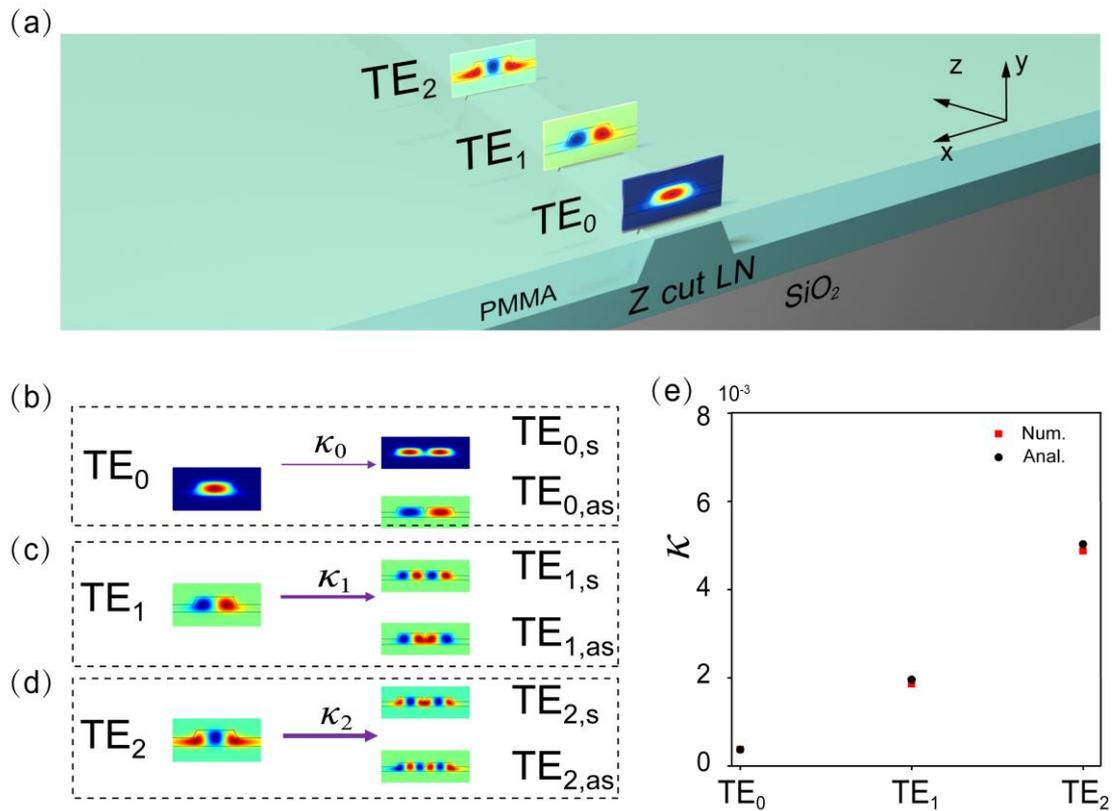

FIG.1. (a) Schematic of TE modes in an LNOI single multimode waveguide excited by the electromagnetic wave at 445 nm. (b-d) Presentation chart of $TE_0$, $TE_1$, and $TE_2$ form symmetric and antisymmetric modes with different propagation constants in the coupling of double waveguide systems, in which arrows with different widths represent different coupling coefficients. (e) Comparison of the mode coupling coefficients obtained by the two methods of calculating coupling coefficients. The results of both calculations are considered similar. Where the red squares represent the coupling coefficients calculated using the numerical method, and the black circles represent the analytical method.

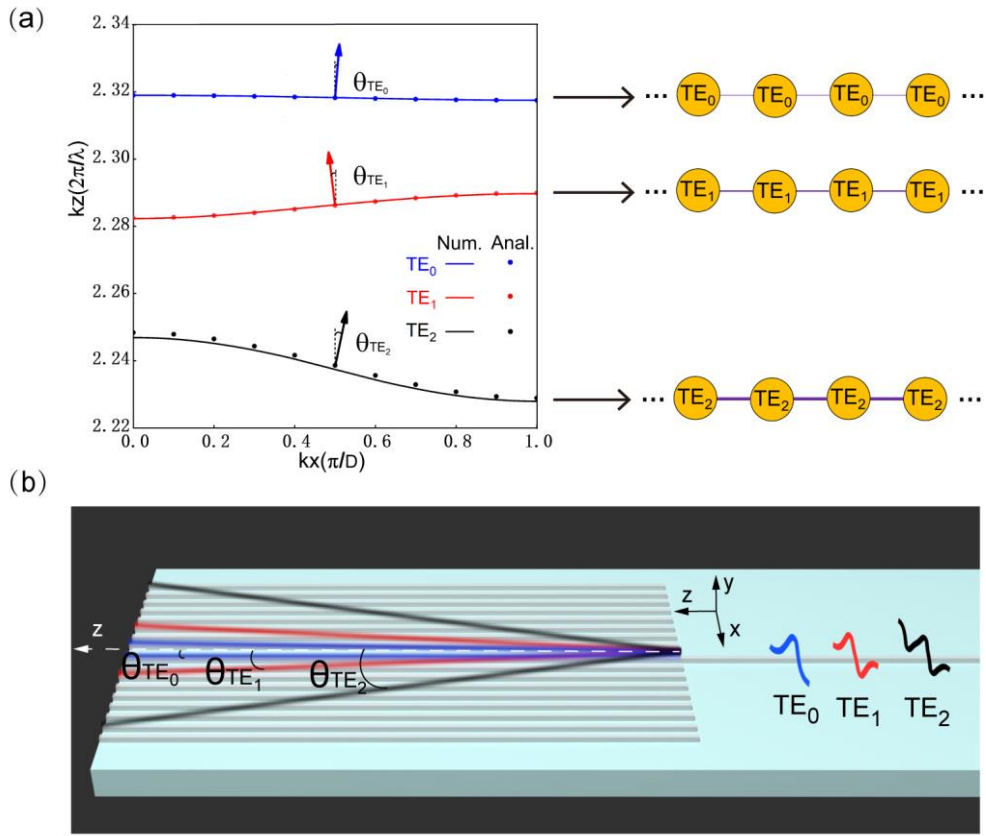

FIG.2. (a) Comparison of the periodic dispersions of three modes obtained from the numerical method and the analytical method. The numerical method is represented by the solid line as shown in the figure. The analytical method of calculating the dispersion can be shown as dotted lines in the figure .where the arrow represents the group velocity dispersion direction (MDA) of each TE mode, and each dispersion corresponds to a one-dimensional atomic chain model. (b) Schematic diagram of transmission of $TE_0$ , $TE_1$ , and $TE_2$ modes in waveguide array.

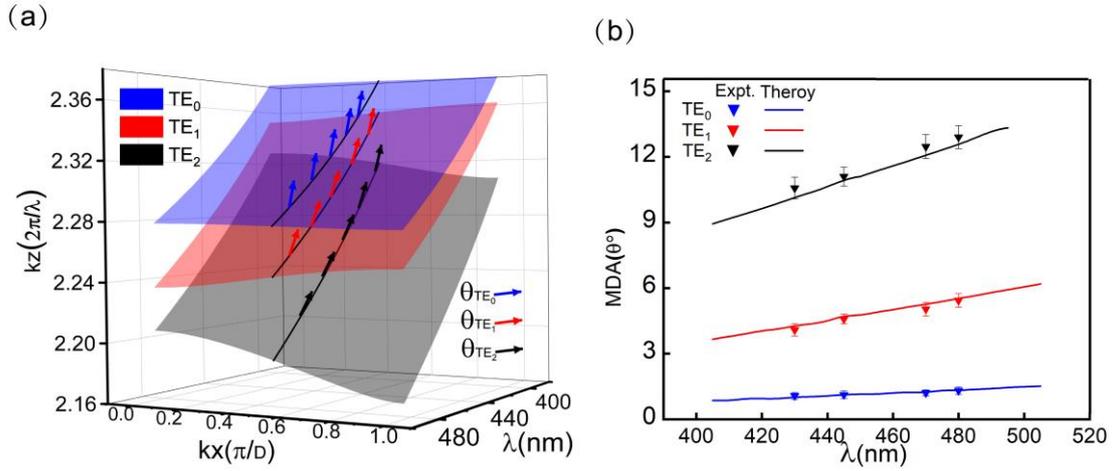

FIG.3. (a) In the wavelength range of 405 nm to 505 nm, the dispersion surface is formed by the coupling of the same TE mode in the waveguide array, and the arrow on the dispersion curve indicates the MDA of the corresponding mode in the waveguide array. (b) The solid line in the figure is the MDA of several TE modes in the wavelength range of 405 nm to 505 nm calculated by COMSOL. The open-label is obtained directly from the experimental data, indicating the MDA of the three TE modes when several wavelengths selected in the experimental test are excited. The ± standard deviation range of the measured data is displayed by the error bar.

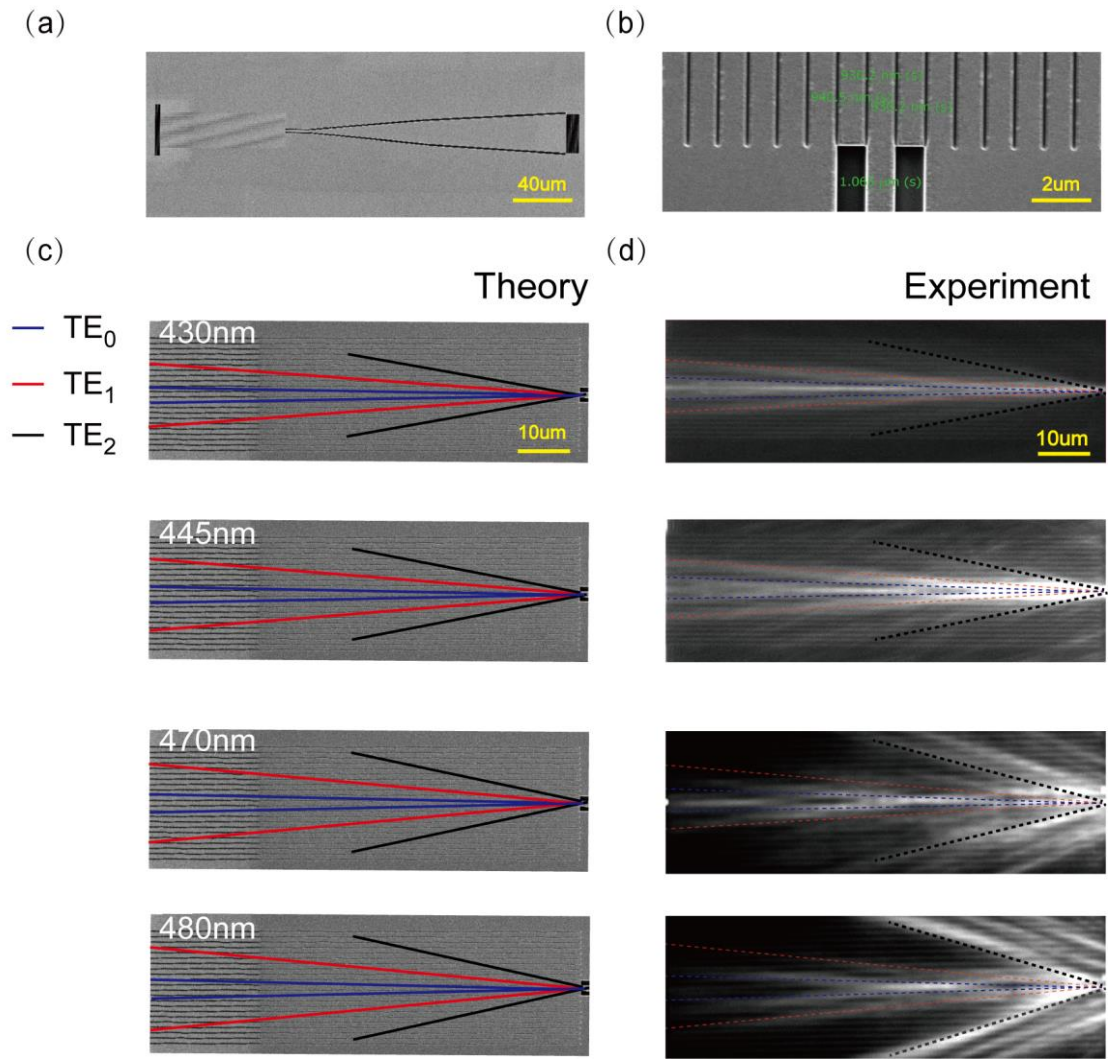

FIG.4. (a) SEM image of lithium niobate waveguide array sample. (b) An enlarged SEM image of the connection between a single multimode waveguide and the waveguide array of the waveguide array sample. (c) The propagation diagram of several TE modes at 430 nm, 445 nm, 470 nm, and 480 nm excitation was calculated by group velocity theory. (d) The transmission diagrams of several TE modes at 430 nm, 445 nm, 470 nm, and 480 nm excitation were measured experimentally.

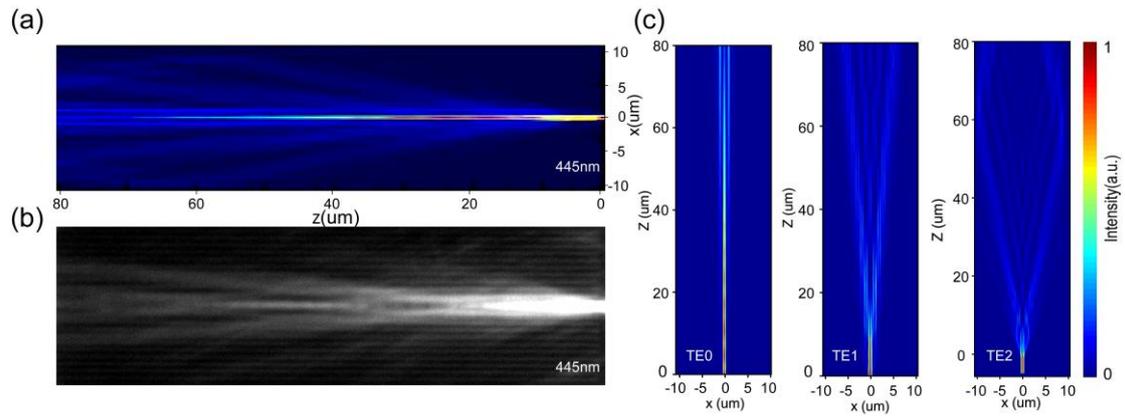

FIG.5. (a) The schematic of mode division demultiplexing for $TE_0$, $TE_1$, and $TE_2$ mode energy ratios of 50%, 25%, and 25%, respectively. (b) Fluorescence trajectory of the mode under 445 nm excitation. (c) The simulation result presents an aerial view of the 80um light field transmission for the three modes.

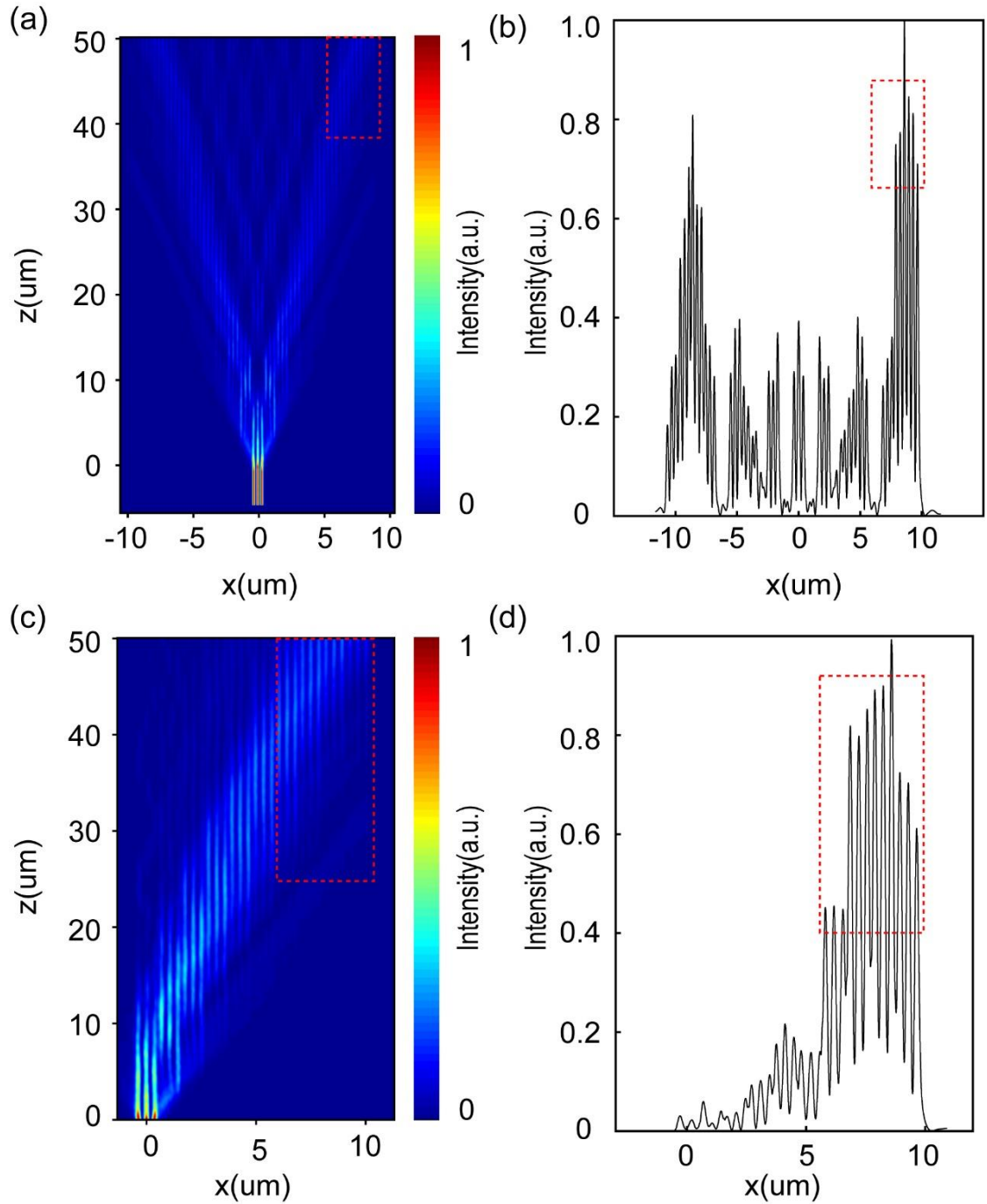

FIG.6. (a) Top view of transmission power when TE$_2$ mode at 445 nm is input from the middle waveguide. (b) Power distribution at the output facet of the waveguide array in Fig.6(a). The red dotted line box in the figure represents the area selected for multiplexing efficiency calculation. Each red dotted line box contains four waveguides. The mode demultiplexing efficiency is defined as the ratio of the output power in the red dashed box region to the input power. The demultiplexing efficiency shown in Fig.

6 (b) is 29%. (c) Top view of transmission power when TE$_2$ mode at 445 nm is input from the boundary waveguide. (d) Power distribution at the output facet of the waveguide array in Fig.6(c). The corresponding multiplexing efficiency in Fig.6(d) is 88.3%.

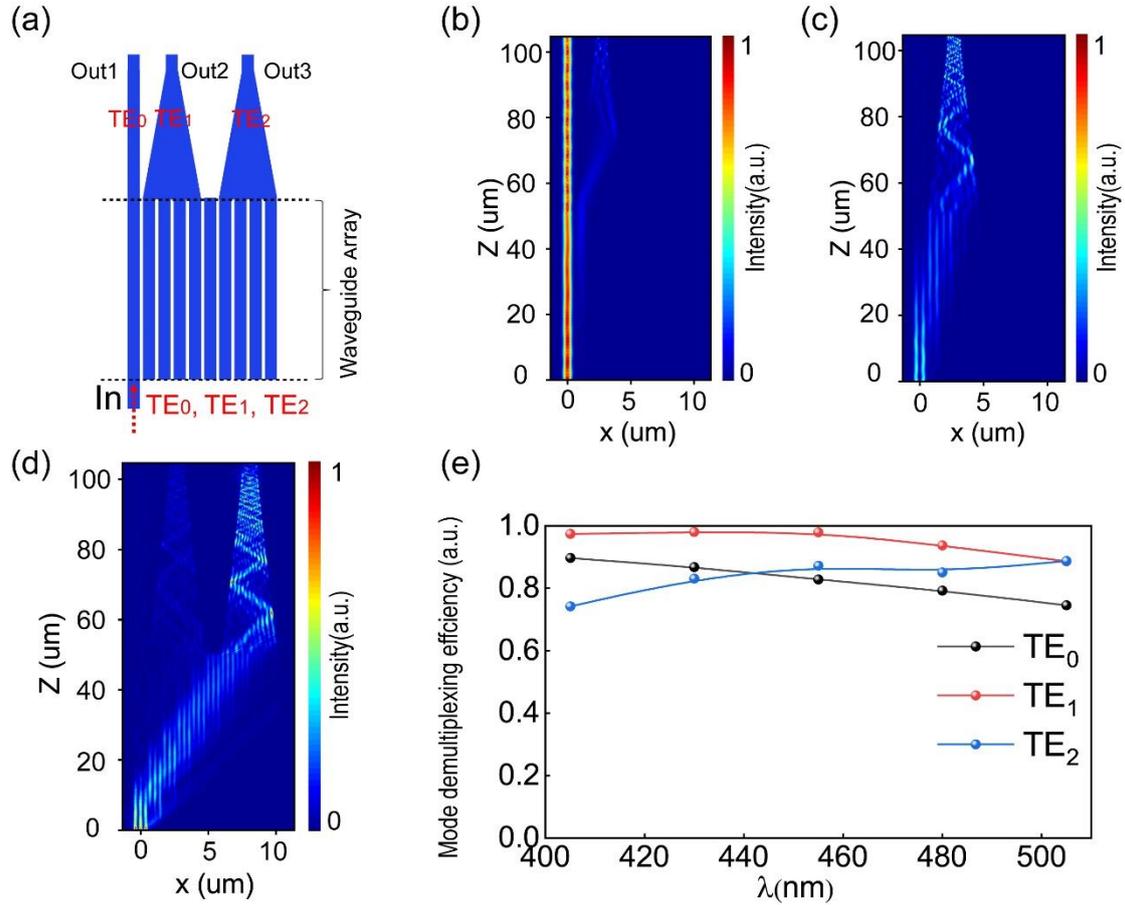

FIG.7. (a) Schematic diagram of a mode division demultiplexer with output ports. (b) Top view of the transmission path of TE$_0$ mode in the demultiplexer when simulating 405 nm excitation. (c) Top view of the transmission path of TE$_1$ mode in the demultiplexer when simulating 405 nm excitation. (d) Top view of the TE$_2$ mode transmission path in the demultiplexer when simulating 405 nm excitation. (e) Schematic of demultiplexing efficiency of TE$_0$, TE$_1$, and TE$_2$ modes in the wavelength range of 405 nm to 505 nm.

|  | 430nm | | 445nm | | 470nm | | 480nm | |
| --- | --- | --- | --- | --- | --- | --- | --- | --- |
|  | Theory | Exp | Theory | Exp | Theory | Exp | Theory | Exp |
| $\theta_{TE0}$ | 1.01° | 1.08° | 1.13° | 1.12° | 1.24° | 1.21° | 1.33° | 1.31° |
| $\theta_{TE1}$ | 4.27° | 4.08° | 4.71° | 4.55° | 5.26° | 5.03° | 5.52° | 5.44° |
| $\theta_{TE2}$ | 10.15° | 10.56° | 11.01° | 11.21° | 12.07° | 12.97° | 12.57° | 13.72 |

TABLE.1. The table shows the theoretically calculated MDA and experimentally tested MDA when the wavelength of the laser is 430 nm, 445 nm, 470 nm, and 480 nm, respectively.

# Supplemental Material for Integrated Broadband Mode Division Demultiplexer in Waveguide Arrays

Yu-le Zhao, Chong Sheng, Zi-yi Liu, Shi-ning Zhu, Hui Liu*

*National Laboratory of Solid State Microstructures, School of Physics, Collaborative Innovation Center of Advanced Microstructures, Nanjing University, Nanjing 210093, China*

*liuhui@nju.edu.cn

**Section I. A method for calculating MDA in periodic dispersion**

In order to calculate the MDA of periodic dispersion, we fit a polynomial to the curve, a fifth-order polynomial can be obtained, and then the derivative function of the polynomial can be obtained: $\frac{d[n_{eff}(k)]}{dk}$. Finally, calculate the maximum absolute value $|\frac{d[n_{eff}(k)]}{dk}|_{max}$ of the derivative function in the range of $0 < k < 1$, then:

$$tan\theta = |\frac{dk_z}{dk_x}|_{max} = \frac{2D}{\lambda}|\frac{d[n_{eff}(k)]}{dk}|_{max} \qquad (S1)$$

$$\theta = arctan(\frac{2D}{\lambda}|\frac{d[n_{ef}(k)]}{dk}|_{max}) \qquad (S2)$$

Therefore: $$\theta_m = arctan(\frac{2D}{\lambda}|\frac{d[n_{eff.m}(k_m)]}{dk_m}|_{max}) \qquad (S3)$$

Where m is the order of the mode, $n_{\text{eff},m}$ is the effective refractive index of the m-order mode, and D is the period of the waveguide arrays. The MDA corresponding to each curve is calculated by the formula (S3), and the angle corresponding to each mode shown in Table 1 can be obtained. Formula (S3) is under the condition of weak coupling.

**Section II. Relationship between MDA and waveguide sidewall inclination angle**

In this part, we mainly use COMSOL software to calculate and analyze the relationship between the MDA and the side wall inclination angle of a single waveguide in the waveguide array. In the calculation, we change the side wall inclination angle of each waveguide in the waveguide array, as shown in Fig.S1a. Then we can calculate the MDA of $TE_0$, $TE_1$, and $TE_2$ modes under each side wall inclination angle, and the calculation results are shown in Fig.S1b. As shown in, the results show that the MDA of the higher-order mode is more sensitive to the change of the side wall inclination angle than the low-order mode, but the change is very small on the whole. The MDA of the most dramatic high-order mode only changes by 4 ° within the change range of the side wall inclination angle from 55 ° to 90 °, which indicates that when the device processing conditions change, such as when the processing current leads to different side wall inclination angles of the waveguide, waveguide array device can still maintain a good mode division demultiplexing function, which shows that waveguide array mode division demultiplexing device is a device with good fabrication tolerance.

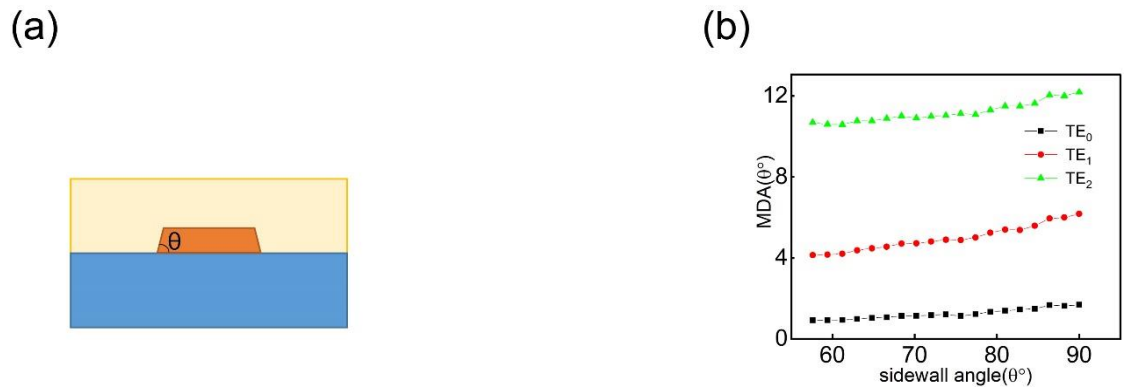

FIG.S1. (a) Schematic diagram of waveguide sidewall inclination angle. (b) The relation curve between the MDA of $TE_0$, $TE_1$, and $TE_2$ modes and the inclination angle of the waveguide sidewall was calculated by COMSOL software.

**Section III. Sample fabrication and measurement**

**Sample fabrication**

1. Preparation of europium doped PMMA solution

We dissolve PMMA particles (Poly Methyl Methacrylate) with toluene solvent to prepare a PMMA solution with a concentration of 15mg/ml, then add 25mg/ml particles containing rare earth europium (Tris(1,3-diphenyl-1,3-propanedionato)(1,10-phenanthroline)europium(III)), and then use a magnetic mixer to stir for 24h to prepare the PMMA solution we need.

2. Fabrication of waveguide arrays

Commercial Z-cut LNOI wafers (NANOLN, Jinan Jingzheng Electronics Co., Ltd.) were used in the sample manufacturing process. First, a 50 nm thick silver film was sputtered on the LNOI wafer by magnetron sputtering. Then, the waveguide array and grating coupler were etched with a focused ion beam (Fei double beam Helios nanolab 600i, 30 keV, 80 PA). The length of the waveguide array is 80 um, a total of 20 waveguides. The depth of the gap between waveguides is 185 nm, and the bottom width of a single waveguide is 1um. Then we immersed the sample in dilute nitric acid to remove the silver film.

3. PMMA solution spin coating

We use the glue homogenizer to spin the configured PMMA solution onto our samples. The glue homogenizer is set with a forward rotation speed of 1000r/min, a time of 6S, and a backward rotation speed of 3000r/min, a time of 30s. After the spin coating is completed, we use 75 ° to dry it on the heating table. The final thickness of the PMMA layer is about 400 nm.

**Experimental measurement**

In the experiment, the laser (690 nm-1040 nm) (Spectra-physics, Mai Tai HP) can obtain the wavelength (345 nm-520 nm) we want for the experiment after the frequency doubling device, and then the 445 nm laser is focused by the microscope objective (Olympus Plan Achromat Objective 20x/0.4), then incident on the grating coupler, and then coupled to the waveguide array. When different modes of light are transmitted in the waveguide array, 610 nm fluorescence can be excited. Before the signal is collected by the camera, we add a 550 nm long wave pass color filter, so that the short wavelength light can be filtered out, so that the fluorescence transmission image can be seen in the camera (Hamamatsu, ORCA-Flash 4.0, C11440-42U). The specific actual optical path diagram is shown in Fig.S2.

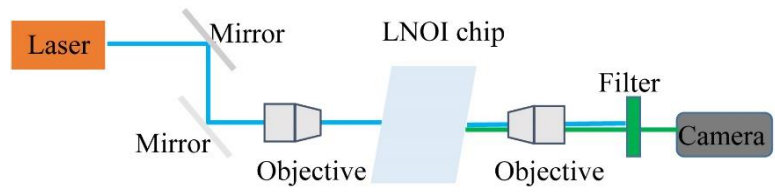

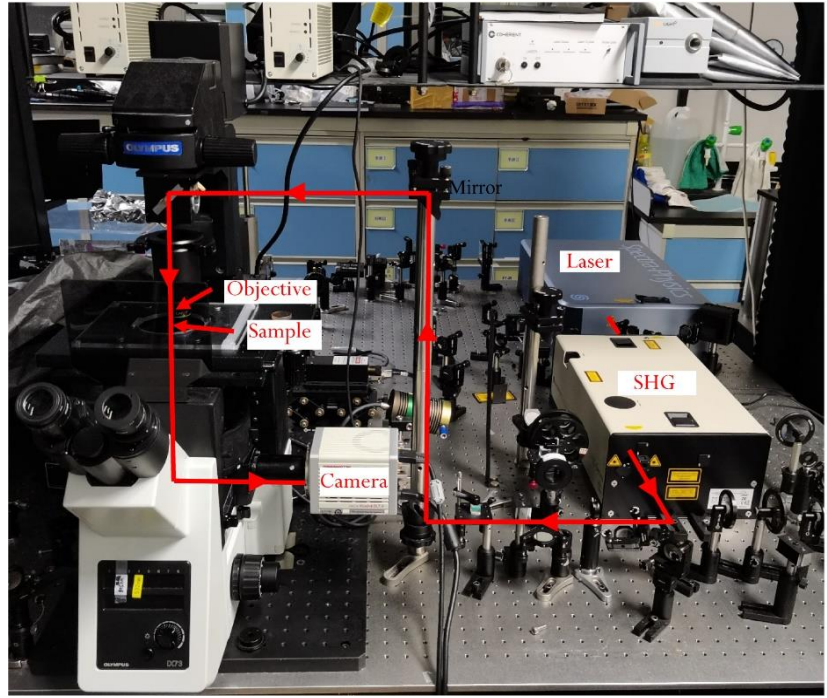

FIG.S2.　Experimental setup